\documentclass[twocolumn,showpacs,aps,prl]{revtex4}
\usepackage{epsfig,amssymb,amsmath}
\usepackage[usenames]{color}
\usepackage{ulem}

\begin{document}

\title{Phase Winding a Two-Component BEC in an Elongated Trap: Experimental
Observation of Moving Magnetic Orders and Dark-bright Solitons}
\author{C. \surname{Hamner}}
\thanks{These authors contributed equally to this work}
\author{Yongping \surname{Zhang}}
\thanks{These authors contributed equally to this work}
\thanks{Present address: Quantum Systems Unit, Okinawa Institute of Science
and Technology, Okinawa 904-0495, Japan}
\author{J.J. \surname{Chang}}
\thanks{These authors contributed equally to this work}
\author{Chuanwei \surname{Zhang}}
\email{chuanwei.zhang@utdallas.edu}
\thanks{Present address: Department of Physics, the University of Texas at
Dallas, Richardson, TX 75080 USA}
\author{P. \surname{Engels}}
\email{engels@wsu.edu}
\affiliation{Washington State University, Department of Physics and Astronomy,
Pullman,Washington 99164, USA}

\begin{abstract}
We experimentally investigate the phase winding dynamics of a harmonically
trapped two-component BEC subject to microwave induced Rabi oscillations
between two pseudospin components. While the single-particle dynamics can be
explained by mapping the system to a two-component Bose-Hubbard model,
nonlinearities due to the interatomic repulsion lead to new effects observed
in the experiments: In the presence of a linear magnetic field gradient, a
qualitatively stable moving magnetic order that is similar to
antiferromagnetic order is observed after critical winding is achieved. We
also demonstrate how the phase winding can be used to generate copious
dark-bright solitons in a two-component BEC, opening the door for new
experimental studies of these nonlinear features.
\end{abstract}

\maketitle

%----------Introduction----------------
Ferromagnetic and antiferromagnetic (AF) orders are two important and
fundamental linear magnetic orders in material physics. For instance, it is
well known that AF order exists in the underdoped and low temperature region
of the phase diagram for high temperature cuprate superconductors \cite{Lee}%
. Ultra-cold atoms provide a clear and highly controllable experimental
platform for emulating various condensed matter phenomena. In ultra-cold
atomic gases, AF order has been predicted to exist for both bosons and
fermions confined in optical lattices, but reaching the required low
temperatures is very difficult \cite{Hulet2012,Esslinger2013}. For cold
atoms confined in optical lattices, AF order corresponds to a quantum state
where atoms at alternating lattice sites have opposite pseudospins and
possess long range phase coherence. For a continuous two-component BEC in a
harmonic trap, there are no such discrete lattice sites, but an AF order can
still be defined similarly to that in lattices. Each spin component contains
periodic and spatially well separated parts and different spin components
appear alternating in space. In lattice system, the lattice periodicity sets
the AF length scale, whence for a continuous system the minimum domain
spacing is limited by the spin healing length. Two-component BECs contain
rich physics and have been investigated extensively in the past decade in
both experiment and theory \cite{Recati}. Notable phenomena include the
analogy to Josephson junction effects for a BEC in a double well potential
\cite{Holland1999,Stenholm1999}, the interaction induced phase separation
\cite{Hall1998,Saito2011}, counterflow induced modulational instability \cite%
{Hoefer}, novel types of solitons \cite{Hoefer}, etc.

\begin{figure}[tbp]
\leavevmode \epsfxsize=3.2in \epsffile{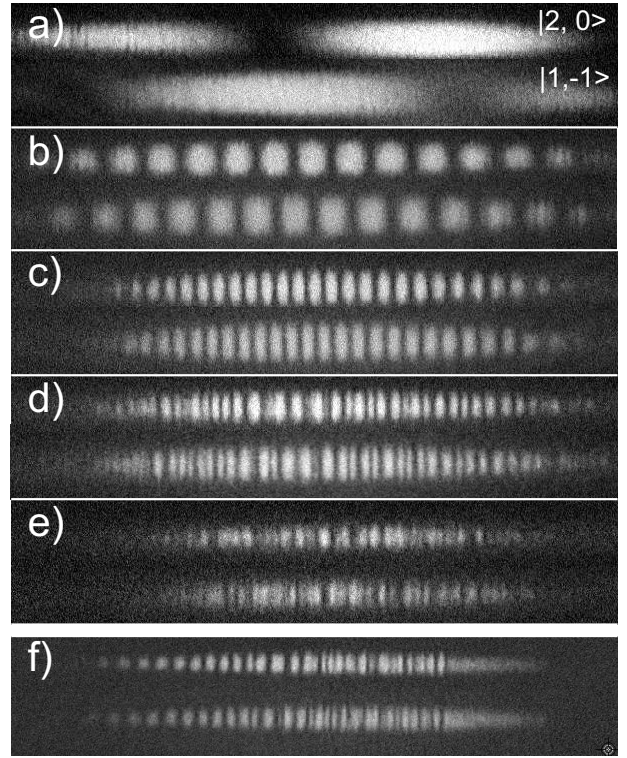}
\caption{Phase winding with a detuning gradient across the axial extent of
the BEC. The $|2,0\rangle$ state (top cloud in each panel) is coupled to the
$|1,-1\rangle$ (bottom cloud) via a microwave pulse of duration a)~10~ms,
b)~100~ms, c)~200~ms, d)~300~ms, and e)~900~ms. f) Image taken at larger
Raman detuning, but similar gradient, and a winding duration of 600~ms.}
\label{winding}
\end{figure}

%----------Description of experiment-----------

In this Letter, we experimentally investigate the dynamics of an
elongated two-component BEC subject to a Rabi coupling between the
two components exposed in the presence of a linear magnetic field
gradient. We show that the dynamics can lead to a phase resembling
AF order. The strong nonlinear interactions in the BEC play a key
role, without them periodic
winding/unwinding cycles analogous to the ones of \cite%
{Cornell1999,Holland2000} are observed instead. We present an insightful
mapping to a Bose-Hubbard model that explains these regular cycles. For the
nonlinear case, our experiment and numerics show a succession of stages
during the winding. First a period of regular winding exists, which is
followed by the emergence of an AF-like pattern, and under the right
conditions a dressed state appears. We also demonstrate how such Rabi
windings can be employed to generate trains of dark-bright solitons.

%----------Description of experiment-----------

To showcase the winding dynamics, we start with a BEC containing about
450,000 $^{87}$Rb atoms in the $|F,m_{F}\rangle $ = $|1,-1\rangle $
hyperfine state. The condensate is confined in an effectively
one-dimensional geometry formed by a 1064~nm single beam optical dipole trap
with measured trap frequencies of $\{\omega _{x},\omega _{y},\omega
_{z}\}=2\pi \cdot \{178,145,1.5\}$~Hz. The atoms are coupled to the $%
|2,0\rangle $ hyperfine state via a microwave pulse of duration $t$. The
scattering lengths are $a_{\uparrow \uparrow }=100.4a_{B}$, $a_{\downarrow
\downarrow }=94.57a_{B}$, and $a_{\uparrow \downarrow }=98.13a_{B}$ \cite%
{Servaas}, where $a_{B}$ is the Bohr radius. The choice of the $|1,-1\rangle
$ and $|2,0\rangle $ state produces a weakly immiscible system as determined
from the mean field condition, $a_{\uparrow\downarrow}^2 >
a_{\uparrow\uparrow}a_{\downarrow\downarrow}$, but similar dynamics are
observed for weakly miscible states as well \cite{OtherStatesFootNote}. The
strength of the coupling pulse is characterized by the Rabi frequency $%
\Omega _{0}$ which we measure to be 7.4~kHz, when on resonance. The atoms
are placed in a magnetic field that consists of a 1~G field in the vertical
(y) direction and a magnetic gradient field along the axial (z) direction
yielding a z-gradient of about 0.017~G/cm. The gradient provides the means
by which the detuning $\delta $, and hence the local Rabi frequency $\Omega $%
, varies across the cloud \cite{ZeemanShiftFootNote}. In our experiments,
unless noted otherwise the detuning at the center of the BEC is chosen to be
1.3~kHz and the point of zero detuning is located to the left of the BEC in
the images. The local Rabi frequency $\Omega_{eff} = \sqrt{%
\Omega_{0}^2+\Delta^2}$ increases as one moves across the BEC to the right.
For imaging we use a spin selective technique described in \cite{Hamner2011}%
. It involves a short expansion period during which no Rabi drive is
applied. Prior to the sudden turn off of the dipole trap for imaging, the
two states are vertically overlapped.

\begin{figure}[tbp]
\center
\includegraphics[width=6.6cm]{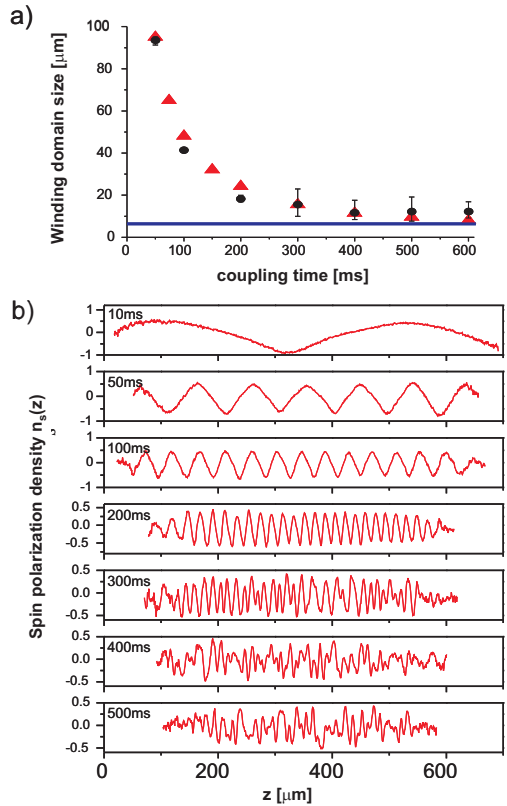}
\caption{(a) Saturation of the spatial spin polarization. The domain size,
measured in the central region of the BEC, stops decreasing after t$\approx$%
300~ms. Overlaid with the experimental results (black circles), where the
error bar indicates the variation in the domain size across the central
region of the BEC, are numerical results (red triangles). The horizontal
line (blue) indicates twice the spin healing length for the initial atom
number of the BEC. Panel b) plots the radially integrated spin polarization
for the winding durations indicated in each panel.}
\label{SpinPlot}
\end{figure}

\begin{figure*}[tbp]
\centering
\centerline{\includegraphics[width=6.5in]{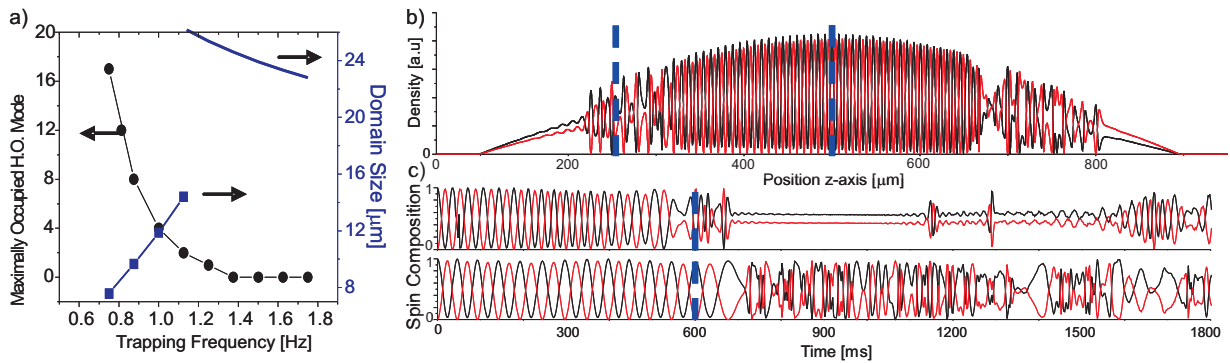}}
\caption{(a) Effect of the axial confinement on the maximally wound single
particle system in the Bose-Hubbard model (black circles). The spin domain
size for these maximal windings (blue squares) and twice the harmonic
oscillator length (blue solid line) are overlayed. (b,c) Numerical results
of the 1D GP equation for experimental parameters. b) Density distribution
for the two spin states after 600~ms of the applied microwave coupling. c)
Temporal evolution of the spin populations at z=250~$\protect\mu $m and
z=500~$\protect\mu$m for the upper and lower panels respectively. In this
graph the chosen 2 ms timesteps leads to aliasing of the fast and regular
Rabi cycles. (color online)}
\label{GPE}
\end{figure*}

While the microwave coupling is applied, windings develop that move in space
with a fixed speed set by the local Rabi frequency \cite{Otago}. The windings move out of
the BEC at one end while new windings emerge at the other end. In Fig.~\ref%
{winding}(a), a short coupling pulse lasting $t=10$~ms creates only two
windings across the cloud. Longer coupling pulses result in more windings
with very regular spacing along the axial direction (Fig.~\ref{winding}
(b,c)). Interestingly, for the chosen parameters this increase in the number
of windings ceases when the winding duration reaches $t\approx300$~ms,
corresponding to an average experimentally observed domain spacing of 15$%
~\mu $m and minimum domain spacing of 8$~\mu$m (Fig.~\ref{winding}(d)).
Following this duration the pattern remains qualitatively unchanged for
several hundred ms, in the sense that experimental images taken during this
interval show domains of similar size, albeit the exact position of the
detected domains varies from shot to shot. This long-time behavior is in
stark contrast to the behavior observed in less elongated trapping
geometries in which the condensate winds and unwinds \cite%
{Cornell1999,Holland2000}. After several hundred ms, atom number losses,
particularly for the $|2,0\rangle $ state, become significant.

To characterize the formation of the magnetic order, we plot the minimum
domain size of each spin component versus the coupling time in Fig.~\ref%
{SpinPlot} a). We see that the experimental domain size decreases and then
saturates after $t\approx 300~ms$. The solid (blue) horizontal line
indicates 2$\zeta _{spin}$ for the initial atom number. For these
experimental parameters, the magnetic order becomes qualitatively fixed when
the domain size approaches twice the spin healing length $\zeta _{spin}=1/%
\sqrt{8\pi n|a|}$, where n is the local density, $a=(2a_{\downarrow \uparrow
}-a_{\downarrow \downarrow }-a_{\uparrow \uparrow })/2$, and $a_{\alpha
\beta }$ are the s-wave scattering lengths \cite{SpinHealingFootNote}.

In Fig.~\ref{SpinPlot} b), we plot the experimentally observed
radially integrated spin polarization density $n_{s}\left( z\right)
=\left( n_{\uparrow }\left( z\right) -n_{\downarrow }\left( z\right)
\right) /\left( n_{\uparrow }\left( z\right) +n_{\downarrow }\left(
z\right) \right) $ after several different winding durations. The
emergence of a magnetic order similar to the AF state, after the
Rabi winding saturates, is clearly evidenced by the periodic
variation of the spin polarization density in this system. However,
$n_{s}\left( z\right) $ does not reach unity after a long time
(i.e., the density of one spin component does not fully disappear).
We attribute this to the large kinetic energy needed for the
complete disappearance of one spin component, to the finite
resolution of the imaging system, or to possible expansion dynamics
during the TOF imaging. Note that in contrast to AF orders in
optical lattices where AF ground states require ultra-low
temperature and entropy, the moving AF orders observed here are
induced by the external strong Rabi coupling, and the temperature of
the BEC plays a negligible role in such dynamics.

For experimental parameters where the spatial variation in the winding rate
is increased, a peculiar and qualitatively different behavior can be
observed: In this case our experimental as well as numerical studies reveal
that extended parts of the BEC can enter a dressed state characterized by
the absence of any winding dynamics in both pseudo-spin components \cite%
{DressedStates}. An example is shown in Fig. \ref{winding}f) where a larger
winding rate was produced by choosing a similar magnetic gradient as before
but a detuning of 4~kHz. After a coupling time of 600~ms the left edge of
the BEC still exhibits Rabi winding while the right edge has evolved into a
dressed state. The two regions are separated by a region of qualitatively
stable AF ordering.

The observed dynamics are well reproduced by numerical simulations using the
one dimensional Gross-Pitaevskii (GP) equation. Choosing the units of
energy, time and length of the system as $\mathbf{\hbar }\omega _{z}$, $%
\omega _{z}^{-1}$, and $\sqrt{\mathbf{\hbar /}m\omega _{z}}$, we can write
the coupled dimensionless equations as%
\begin{equation}
i\frac{\partial }{\partial t}%
\begin{pmatrix}
\Phi _{\uparrow } \\
\Phi _{\downarrow }%
\end{pmatrix}%
=%
\begin{pmatrix}
H_{\uparrow }+\Delta (z) & \Omega _{0} \\
\Omega _{0} & H_{\downarrow }%
\end{pmatrix}%
\begin{pmatrix}
\Phi _{\uparrow } \\
\Phi _{\downarrow }%
\end{pmatrix}%
.  \label{G-P}
\end{equation}%
Here, $\Phi _{\uparrow }$ and $\Phi _{\downarrow }$ represent condensate
wavefunctions in the hyperfine states $|1,-1\rangle $, and $|2,0\rangle $
respectively. $H_{\uparrow }=H_{0}+2N\sqrt{m\omega _{x}\omega _{y}/\hbar
\omega _{z}}(a_{\uparrow \uparrow }|\Phi _{\uparrow }|^{2}+a_{\uparrow
\downarrow }|\Phi _{\downarrow }|^{2})$, and $H_{\downarrow }=H_{0}+2N\sqrt{
m\omega _{x}\omega _{y}/\hbar \omega _{z}}(a_{\downarrow \downarrow }|\Phi
_{\downarrow }|^{2}+a_{\uparrow \downarrow }|\Phi _{\uparrow }|^{2})$, where
$H_{0}=-\frac{\partial ^{2}}{2\partial z^{2}}+\frac{z^{2}}{2}$, and the
second terms describe the mean field nonlinear interaction between atoms. $N$
is total atom number. $\Delta (z)=\Delta _{0}+\delta z$ includes a constant
detuning $\Delta _{0}$ and the detuning $\delta$ caused by the magnetic
gradient. We numerically solve the GP equation (\ref{G-P}) using the
experimental parameters as in Fig.~\ref{winding}(a-e) and \ref{SpinPlot},
and determine the domain spacing after various durations of the microwave
pulse (red triangles in Fig.~\ref{SpinPlot} a)). We find a good agreement
between the numerical results and the experimental data shown. Our numerical
simulations also reveal that the spin density polarization $n_{s}\left(
z\right) $ does not reach unity after a long duration in the AF phase,
agreeing with the experimental observation.

To emphasize the importance of the nonlinear interactions for these
dynamics, we contrast the observed behavior with the predictions of
the a single-particle picture which ignores meanfield contributions
to the spin dynamics. In this case regular winding/unwinding
processes exist, which can be understood from an interesting and
insightful mapping to a two-component Bose Hubbard model. We start
from the linear system of Eq.(1) by neglecting nonlinear terms and
expand $\Phi _{\uparrow }$ and $\Phi _{\downarrow }$ using the
harmonic oscillator basis $\Psi _{j}$, $\Phi _{\uparrow
}(z)=\sum_{\substack{ j}} a_{j}\Psi _{j}(z)$ and $\Phi _{\downarrow
}(z)=\sum_{\substack{ j}} b_{j}\Psi _{j}(z)$. Substituting these
expansions into Eq.(1), we obtain

%%%%%%%%%%%%%%
\begin{align}
& i\frac{\partial a_{j}}{\partial t}=\frac{1}{2}ja_{j}+\Omega
_{0}b_{j}+\Delta _{0}a_{j}+\delta \sqrt{\frac{j+1}{2}}a_{j+1}+\delta \sqrt{%
\frac{j}{2}}a_{j-1},  \notag \\
& i\frac{\partial b_{j}}{\partial t}=\frac{1}{2}j b_{j}+\Omega _{0}a_{j},
\end{align}
which is reminiscent of a two-component Bose-Hubbard model. This model can
be interpreted as a lattice system subject to a linear potential $\frac{1}{2}%
z$, leading to the on-site energy $j/2$. The linear potential prevents the
atom from climbing to large $j$. On the other hand, the effective tunneling
coefficient is anisotropic: it is $\delta \sqrt{(j+1)/2}$ for $j\rightarrow
j+1$, but $\delta \sqrt{j/2}$ for $j\rightarrow j-1$. Therefore the atoms
prefer to tunnel to large $j$ sites. The competition between the linear
potential and the anisotropic tunneling leads to a maximum $j_{0}$.
Numerical simulations of Eq.(2) confirm that there exists a maximally
occupied $j$ for the dynamical oscillation of $a_{j}$ and $b_{j}$ with the
initial condition $a_{0}=1$ and $b_{0}=0$. The $j$-th harmonic oscillator
wavefunction has $j$ nodes, yielding $j+1$ possible domains in the density
of each component. The maximally occupied mode achieved during the single
particle winding is plotted in Fig.~\ref{GPE} a) for various axial
confinements. This maximal winding number increases as the axial trap
frequency $\omega _{z}$ is decreased while the the magnetic gradient is kept
fixed.

The addition of the nonlinearity leads to the coupling of atoms to
sites with higher $j$. This is expected based on the following:
First, the repulsive interactions lead to a larger spatial extent of
the BEC, so that larger modes have to be occupied to reach the same
domain spacing. Second, larger interactions reduce $\zeta _{spin}$,
decreasing the minimum possible domain size. The winding dynamics
depart from the single particle like recursions when the system
attempts to flip its order parameter, to begin unwinding.
Fig.~\ref{GPE}(b,c) shows results of the 1D GP simulation for the
experimental parameters. Fig.~\ref{GPE} b) shows the density profile
along the z axis after 600~ms winding, i.e. just before the winding
fully saturates. Fig.~\ref{GPE} c) shows time evolution plots for
the local spin composition at the spatial locations indicated by the
vertical lines in b). The deviations from the single particle like
winding occur at different evolution times across the spatial extent
of the BEC and the dynamics across the full extent of the BEC no
longer seem synchronized leading to the AF like ordering. We note
that in the limit of small nonlinearity the numerics recover the
winding/unwinding behavior.

\begin{figure}[tbp]
\leavevmode \epsfxsize=3.2in \epsffile{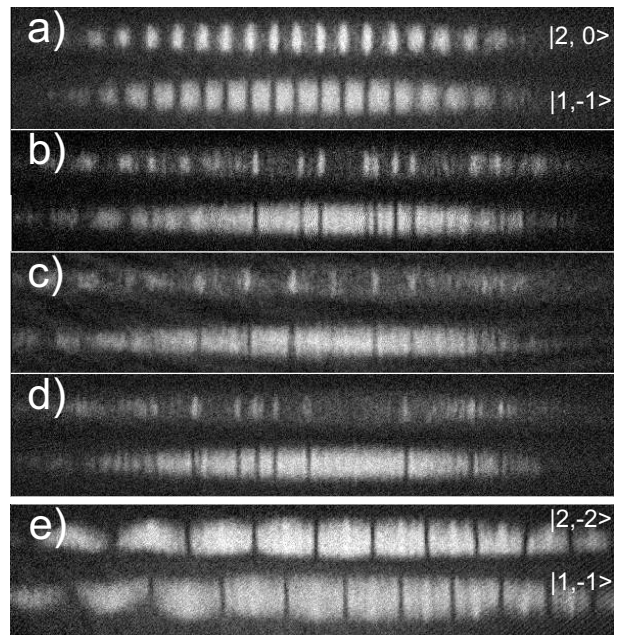}
\caption{Generation of dark-bright solitons via phase winding. After
applying a coupling pulse of $t=200$~ms, the magnetic gradient and coupling
are jumped off and the clouds are allowed to evolve in-trap for a)~100~ms,
b)~300~ms, c)~400~ms and d)~500~ms before imaging. (e) A BEC of atoms in the
$|1,-1\rangle$ and $|2,-2\rangle$ state is wound into a soliton train and
let evolve for 60~ms. Here a DB soliton train with alternating polarity is
generated.}
\label{dbsolis}
\end{figure}

The phase winding can also be exploited as a tool to generate copious
dark-bright solitons, enabling studies of their dynamics. During the phase
winding, the quantum mechanical phase advances by $\pi$ between consecutive
domains of the same component (neglecting a minor deviation from $\pi$
induced by the finite and spatially varying detuning of the Rabi drive).
This is a natural starting point for the generation of dark-bright solitons
in which a dark soliton, with a phase jump of $\pi$, in one component is
filled by a bright soliton in the second component \cite{Anderson2001}. As
we experimentally demonstrate, the phase winding pattern can be transformed
into a dark-bright soliton train. For the data shown in Fig.~\ref{dbsolis},
we start by reproducing the situation in Fig.~\ref{winding} (c) and then
abruptly turn off the applied gradient as well as the Rabi drive. The
subsequent in-trap evolution is strongly influenced by the difference in
lifetime between the $|1,-1\rangle $ and $|2,0\rangle $ states (30 sec vs
0.5 sec respectively). As atoms are preferentially lost from the $%
|2,0\rangle $ state, the domains in the $|1,-1\rangle $ state approach each
other. Assisted by the phase difference between domains in the $|1,-1\rangle
$ state, dark solitons filled by atoms of the $|2,0\rangle $ state form.
Figs.~\ref{dbsolis}(b)-(d) show the evolution of these dark-bright (DB)
solitons. The consistent width and long lifetime of these structures provide
evidence for their solitonic character as seen under other experimental
conditions \cite{Sengstock2008,Hamner2011,PanosEngels}. Though many DB
solitons are formed that are initially equally spaced, the regularity of the
arrangement is lost after longer wait times. A similar evolution of the
wound BEC into solitons can also be observed for the longer lived states $%
|1,-1\rangle $ and $|2,-2\rangle $ as seen in Fig. ~\ref{dbsolis} (e) with a
60~ms evolution time after the end of the Rabi drive. Here DB solitons form
with alternating polarity in both spin components at the interfaces between
the spin domains. To our knowledge, this constitutes the first generation of
a train of alternatingly polarized DB solitons in a BEC and offers exciting
prospects for the investigation of complex soliton interactions, e.g. the
observation of interesting "soliton molecules" \cite{SolitonClusters}. The
ability to generate large numbers of dark-bright solitons with well defined
initial periodicity may also be an effective starting point for
investigations of soliton gases \cite{SolitonGas1,SolitonGas}.

\textbf{Acknowledgments}: C.H., J.J.C., and P.E. acknowledge financial
support from NSF and ARO. Y.Z. and C. Z. acknowledge support from ARO
(W911NF-12-1-0334), AFOSR (FA9550-11-1-0313), and NSF-PHY (1249293).


\begin{thebibliography}{99}
\bibitem{OtherStatesFootNote} We have experimentally observed similar
winding dynamics and soliton generation in the longer lived states $%
|1,-1\rangle$ and $|2,-2\rangle$ which are weakly miscible with scattering
length $a_{\downarrow\downarrow}=100.4a_0$, $a_{\uparrow\uparrow}=98.98a_0$,
and $a_{\downarrow\uparrow}=98.98a_0$.
\bibitem{ZeemanShiftFootNote} In $^{87}$Rb, the linear Zeeman effect leads
to a shift of approximately 700~kHz/Gauss of the $|1, -1\rangle$ state,
while the $|2, 0\rangle$ state remains unaffected (to first order). The
variation of the detuning across the BEC also leads to a variation of the
amplitude of the Rabi oscillation, which, however, is negligible for the
chosen parameters.
\bibitem{SpinHealingFootNote} For our initial atom number the spin healing
length at the BEC center is approximately 3.2$~\mu$m.
\end{thebibliography}
\end{document}